\documentclass[aps,pra,reprint,superscriptaddress,secnumarabic]{revtex4-2}

\usepackage{orcidlink} 
\usepackage{graphicx}
\usepackage{dcolumn}
\usepackage{bm}

\usepackage{placeins}

\usepackage{multibib}

\usepackage{braket}
\usepackage{amsmath}
\usepackage{bibentry}

\usepackage{color,soul}
\graphicspath{{figures/}}

\newcommand{\nocontentsline}[3]{}
\newcommand{\tocless}[3]{\bgroup\let\addcontentsline=\nocontentsline#1{#3}\egroup}

\begin{document}

\title{Observation of counterflow superfluidity in a two-component Mott insulator}

%
\author{Yong-Guang Zheng\orcidlink{0000-0002-2724-7340}}

\thanks{Y.-G. Z. and A. L. contributed equally to this work.}

\affiliation{Hefei National Research Center for Physical Sciences at the Microscale and School of Physical Sciences, University of Science and Technology of China, Hefei 230026, China}
\affiliation{CAS Center for Excellence in Quantum Information and Quantum Physics, University of Science and Technology of China, Hefei 230026, China}
\affiliation{Hefei National Laboratory, University of Science and Technology of China, Hefei 230088, China}

\author{An Luo}
\thanks{Y.-G. Z. and A. L. contributed equally to this work.}

\author{Ying-Chao Shen}


\author{Ming-Gen He}


\author{Zi-Hang Zhu}


\author{Ying Liu}


\author{Wei-Yong Zhang}


\author{Hui Sun}

\affiliation{Hefei National Research Center for Physical Sciences at the Microscale and School of Physical Sciences, University of Science and Technology of China, Hefei 230026, China}
\affiliation{CAS Center for Excellence in Quantum Information and Quantum Physics, University of Science and Technology of China, Hefei 230026, China}

\author{Youjin Deng}


\author{Zhen-Sheng Yuan}


\author{Jian-Wei Pan}

\affiliation{Hefei National Research Center for Physical Sciences at the Microscale and School of Physical Sciences, University of Science and Technology of China, Hefei 230026, China}
\affiliation{CAS Center for Excellence in Quantum Information and Quantum Physics, University of Science and Technology of China, Hefei 230026, China}
\affiliation{Hefei National Laboratory, University of Science and Technology of China, Hefei 230088, China}


\date{\today}

\begin{abstract}
The counterflow superfluidity (CSF) was predicted two decades ago.
Counterintuitively, while both components in the CSF have fluidity,
their correlated counterflow currents cancel out leading the overall system to an incompressible Mott insulator.
However, realizing and identifying the CSF remain challenging due to the request on extreme experimental capabilities in a single setup.
Here, we observe the CSF in a binary Bose mixture in optical lattices. 
We prepare a low-entropy spin-Mott state by conveying and merging two spin-1/2 bosonic atoms at every site
and drive it adiabatically to the CSF at $\sim$1 nK.
Antipair correlations of the CSF are probed though a site- and spin-resolved quantum gas microscope in both real and momentum spaces.
These techniques and observations provide accessibility to the symmetry-protected topological quantum matters.
\end{abstract}


\maketitle


\noindent \textbf{Introduction.}
Superfluidity and superconductivity, possessing zero resistance to mass and charge flows respectively,
have attracted intensive interest from both fundamental and applied researches \cite{Feynman1957,Yang1962,Leggett1999}.
The former, superfluidity, discovered in helium \cite{Kapitza1938,Allen1938,Bardeen1957,Osheroff1972}, was realized with quantum gases of bosonic \cite{Anderson1995,Davis1995} and fermionic atoms \cite{Regal2004}. 
Loading the superfluid into an optical lattice demonstrated the quantum phase transition from a superfluid to a Mott insulator \cite{Greiner2002}. 
It was predicted in theory, counterintuitively, merging two Mott insulators (each filled by one of the two components of bosonic atoms) can form an exotic phase of the so-called counterflow superfluidity (CSF), where both components become mobile and exhibit exactly correlated counterflows however the overall Mott insulator of total charges is still maintained \cite{Kuklov2003,Altman2003,Kuklov2004}.

\begin{figure*}
\includegraphics[width=0.99\linewidth]{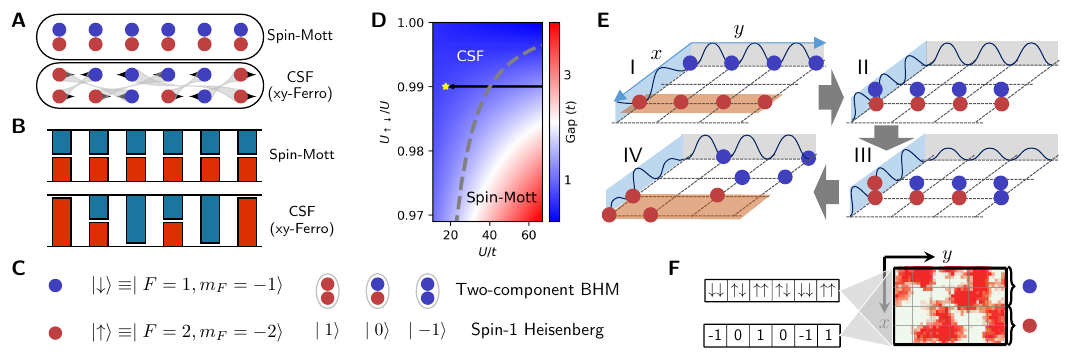}%
\caption{\label{fig:setup}\textbf{Probing counterflow superfluidity in optical lattices.} 
(\textbf{A}) In a two-component Bose mixture, there are two phases of the so-called counterflow superfluid (CSF) and spin-Mott. The correlated CSF of the two components suppresses the total number fluctuations (\textbf{B}), in stark contrast to the conventional superfluid of the single component BHM. 
(\textbf{C}) We encode the two components in the two hyperfine states of the ${}^{87}$Rb atoms (left panel). State mapping between the two-component BHM and spin-1 Heisenberg model (right panel). (\textbf{D}) Phase diagram of the two-component BHM. The dashed line indicates the phase boundary between the CSF and spin-Mott obtained from a mean field theory. The solid arrow is the adiabatic passage to the CSF phase (yellow star) used in our experiments. (\textbf{E}) Illustration of the experimental sequence. I, II. State preparation of the doublon occupied spin-Mott chains; III. Adiabatic passage to the CSF phase; IV. Stern-Gerlach separation and doublon-splitter operations for spin- and number-resolved detection. (\textbf{F}) A typical raw picture (right panel) of the $L=6$ chain after Stern-Gerlach separation and doublon-splitter operations and the corresponding readout in the two-component BHM and Heisenberg model (left panel).}
\end{figure*}

It is proposed that this interesting CSF phase can be examined in terms of pseudo-spins and 
re-interpreted as the xy-ferromagnet of a spin-1 Heisenberg model, 
offering a platform to study quantum magnets \cite{Kuklov2003,Altman2003,Duan2003,Hubener2009,Powell2009,Schachenmayer2015,DeParny2020,Basak2021}. 
When the imbalance between the inter- and intra-component interactions is further enhanced,
the mobility of each spin component disappears, leading to a spin-Mott insulator (Fig. \ref{fig:setup} B). 
The quantum phase transition between the CSF and spin-Mott insulator 
is akin to the superfluid to insulator transition but happens in the spin domain \cite{Schachenmayer2015,DeParny2020}. 

Though great efforts have been devoted both theoretically \cite{Hu2009,Menotti2010,Hu2011,Schachenmayer2015,DeParny2020,Venegas-Gomez2020,Venegas-Gomez2020b,Basak2021} and experimentally \cite{Chung2021,deHond2022,Dimitrova2023} in searching for the CSF, it remains elusive for an experimental demonstration. In principle, this highly-correlated phase could be approached adiabatically from a spin-Mott insulator which is a gap-protected product state and is feasible to realize in experiment \cite{deHond2022}. However, the CSF emerges only below the temperature defined by the superexchange energy $J/k_B$, typically in the order of nanokelvin or even lower in optical lattices. Therefore, challenges arise from two aspects: the preparation of a defect-free spin-Mott insulator and the adiabatic passage to the CSF with negligible heating. The request on cutting-edge techniques of probing the coherence and correlations puts further burden in experimental realization. 

In this work, we report on realizing and indentifying the CSF with a two-component bosonic mixture in optical lattices. We first prepare an ultra-low entropy unity-filling Mott insulator of a single component via exploiting the staggered-immersion cooling \cite{Yang2020}. 
Then an equal mixture of two components is realized by site-resolved spin-flip with high fidelity \cite{Zheng2022}. Further, we develop a method to convey one chain of insulator filled by a single component to its neighboring chain filled by the second component and merge them together via phase-tunable superlattices, creating homogeneous defect-free spin-Mott chains (See SM). Tuning coherently the imbalance between the inter- and intra-component interactions, we adiabatically drive the system to the CSF phase. To detect this phase, we perform Stern-Gerlach separation and doublon-splitter operation (Fig. \ref{fig:setup} F) for a full read-out of the local spin and charge state simultaneously with the quantum gas microscope, which allows us to obtain the correlations of the CSF in both real and momentum spaces.
\\

\noindent \textbf{The CSF in a two-component BHM.}
For simplicity and without loss of generality, we focus on the two-component BHM in one-dimension, which describes our experiments as well
\begin{equation}
\label{equ:TBHM}
 \begin{split}
 \mathcal{\hat{H}}_\textsc{TBH} &=  -t \sum_{i,\sigma} \left(\hat{b}_{i,\sigma}^{\dagger}\hat{b}_{i+1,\sigma}+H.c.\right) \\
  + &\frac{U}{2} \sum_{i,\sigma} \hat{n}_{i,\sigma}\left(\hat{n}_{i,\sigma}-1\right) + U_{\uparrow \downarrow} \sum_i \hat{n}_{i,\uparrow} \hat{n}_{i,\downarrow},
\end{split}
\end{equation}
where $t$ is the tunnelling amplitude, $U$ and $U_{\uparrow \downarrow}$
are respectively the intra- and inter-component on-site interaction,  
$\hat{b}^{\dagger}$ ($\hat{b}$) is the creation (annihilation) operator of bosons, and $\hat{n}_{\uparrow}$ ( $\hat{n}_{\downarrow}$) is the number operator of bosons in the $\ket{\uparrow}$ ($\ket{\downarrow}$) state. 
The footnote $i$ specifies the lattice site in range from 0 to $L-1$ and $\sigma$ denotes the spin. 

For the balanced unity-filling for both components, i.e., $N_{\uparrow}/L=N_{\downarrow}/L=1$, the system first undergoes 
the superfluid-Mott-insulator phase transition 
as $U/t$ is increased, similar to the single-component BHM \cite{Greiner2002}. 
In the Mott regime where $U/t\gg1$ and the superexchange coupling $J=4t^2/U_{\uparrow\downarrow}>u/4$, with $u=U-U_{\uparrow \downarrow}$
the ground state is the CSF. 
If one further increases the interaction such that $J/u\ll1$, 
the ground state becomes a spin-Mott insulator 
with vanishing charge and spin fluctuations \cite{Kuklov2003,Altman2003,Hu2009,DeParny2020}.

In the Mott regime, the two-component BHM within the subspace of 
doublon occupations could be mapped on an effective spin-1 Heisenberg model (See SM).
For $u>0$, the easy-plane anisotropy favors
the ferromagnetism in the $xy$ plane \cite{Kuklov2003,Altman2003,Duan2003,Hubener2009,Powell2009,Schachenmayer2015,DeParny2020,Basak2021,Prufer2022,Chen2023,Feng2023}.
Namely, the CSF phase can be regarded as an xy-ferromagnet.
\\

\noindent \textbf{Probing the CSF in optical lattices.}
We prepare  ${}^{87}$Rb atoms of two internal states of $\ket{\downarrow}\equiv\ket{F=1,m_F=-1}$ and $\ket{\uparrow}\equiv\ket{F=2,m_F=-2}$ in the lowest bands of the optical lattices. A positive anisotropy of $u/U\sim 1\%$ is fixed by the scattering lengths, and the relative $J/u$ is tuned via ramping the lattice depth \cite{Chung2021} to implement the adiabatic passage. 
We create several 1D chains of $\ket{\downarrow}$ atoms along the $y$ direction, which are unity-filling Mott insulators cooled by the staggered superlattice in the $x$ direction. Besides, the chains are cut to a fixed length of $L=6$ by site-resolved addressing \cite{Zheng2022}. Then atoms in every second chain are addressed and flipped to $\ket{\uparrow}$ state, and are further conveyed to merge with the neighboring chains of $\ket{\downarrow}$ atoms by tuning the phase of the $x$ long lattice (Fig. \ref{fig:setup} E). During the conveying of $\ket{\downarrow}$ atoms, the $\ket{\uparrow}$ atoms are pinned by deep traps induced by the light beam of a magic wavelength projected by the digital micromirror device (DMD) (See SM). Now we get several chains occupied by pairs of $\ket{\uparrow}$ and $\ket{\downarrow}$ atoms in single sites, which is the spin-Mott state, akin to the band insulator of fermions. 

\begin{figure}[t!]
\includegraphics[width=0.99\linewidth]{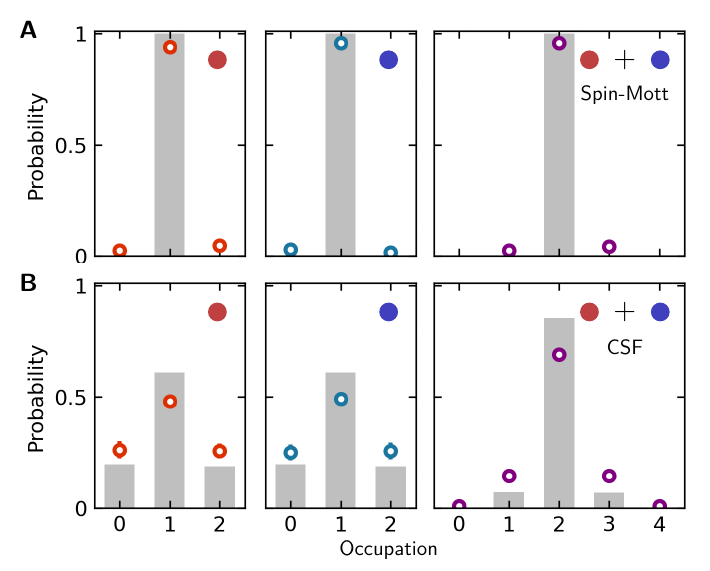}%
\caption{\label{fig:Var}\textbf{Number-squeezing of the total atoms in the CSF.} (\textbf{A}) and (\textbf{B}) show the histograms of the atom occupation in single sites in the spin-Mott and the CSF phase, respectively. 
Left, middle and right panels are for the $\ket{\uparrow}$, $\ket{\downarrow}$ and both atoms, respectively. 
Error bars denote the SEM and are smaller than the markers if not visible. 
Gray bars are predictions from the ground states in the two phases.}
\end{figure}

\begin{figure}[t!]
\includegraphics[width=0.99\linewidth]{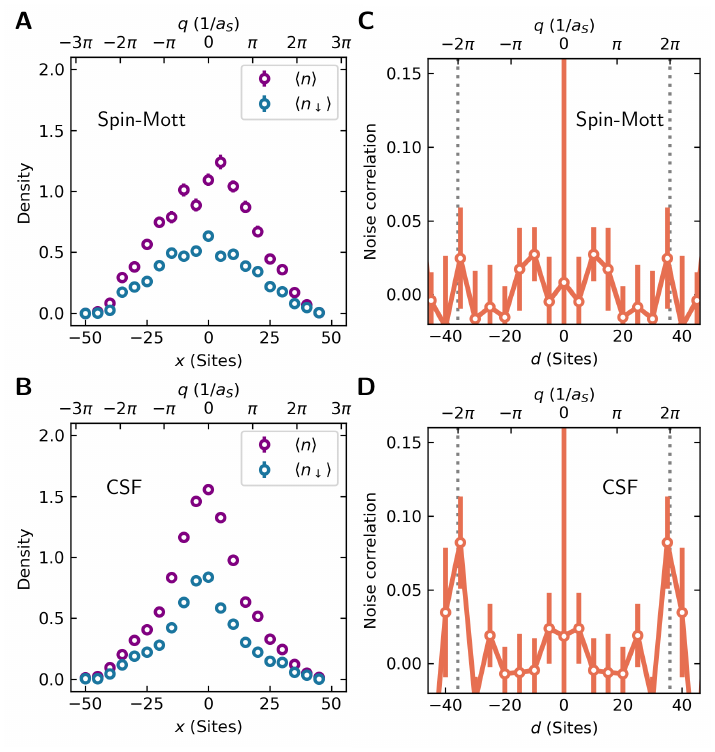}%
\caption{\label{fig:tof}\textbf{Antipair correlations of the CSF.} Density distribution after the time-of-flight (TOF)  in the spin-Mott (\textbf{A}) and CSF phases (\textbf{B}). The purple circles are for the density of the total atoms and the blue circles are for 
the density of the $\ket{\downarrow}$ atoms. 
We further extract the noise correlations from the TOF images 
for the spin-Mott (\textbf{C}) and the CSF (\textbf{D}) phases.
The orange circles are the correlation between the two components.
The associated solid lines are guides to the eye. Error bars denote the SEM and are hidden if not visible. The large error bars in $d=0$ are attributed to the local number fluctuations of the expanded cloud. Vertical dotted lines indicate the predicted positions of the $\pm 2\pi/a_S$ peaks in \textbf{C} and \textbf{D}.}
\end{figure}

The depth of the $x$ short lattice is around 102$E_r$, where $E_r=h^2/(8 m_{\mathrm{Rb}} a_S^2)$ is the recoil energy of the short lattice, $h$ is Planck's constant, $m_{\mathrm{Rb}}$ is the mass of ${}^{87}$Rb atoms, and $a_S=630$ nm is the short lattice spacing. The depth of the $y$ short lattice is 18.4$E_r$. With these parameters, the corresponding intra-component interaction is $U/h=1.2$ kHz, $u/h=12$ Hz and the tunnelling is $t/h=4.9$ Hz. Afterwards, we ramp down the $y$ short lattice adiabatically to 7.4$E_r$, where the tunnelling strength is enhanced to $t/h\sim 52$ Hz and the anisotropy is $u/h=9.1$ Hz. There is still negligible fluctuation of the atom number in each site due to the strong interaction of $U/t =17.5$ but the CSF emerges from the spin exchange process with $J/h=21.5$ Hz. 

Finally, we freeze the system by a quick jump of the depth of the $x$ and $y$ lattices. Before the fluorescence imaging,
we perform Stern-Gerlach separation for spin-resolved readout. This process is similar to the preparation of the initial states of spin-Mott doublons but in the reverse order. In short, the doublons are loaded to the $x$ long lattice and are pinned by a magic wavelength addressing beam that can only be seen by the $\ket{\uparrow}$ atoms. Then the phase of the long lattice is tuned and the $\ket{\downarrow}$ atoms will follow the conveyor belt. The $\ket{\downarrow}$ atoms are conveyed by two short lattice sites in the $x$ direction and the $\ket{\uparrow}$ atoms stay where they are.
Furthermore, to avoid pairwise loss of doublons with identical spins in single sites during the imaging, we split the doublons in the balanced double-wells imposed by the superlattice before detection (Fig. \ref{fig:setup} E, F). Now the doublons are mapped to four lattice sites so we could fully read out the state of the doublons.

We directly probe the occupations in single lattice sites. As shown in Fig. \ref{fig:Var}, while the distributions of the total atom number 
are narrow in both phases, as expected for the strong on-site interaction,
the distributions of the $\ket{\uparrow}$ or $\ket{\downarrow}$ 
are much broader in the CSF phase than in the spin-Mott phase.
The covariance of the two components is further calculated,
which is ${\rm Cov}(\uparrow,\downarrow)=-0.31(1)$ in the CSF, 
indicating the existence of strong correlations 
between the two components. In contrast,  
the spin-Mott phase has ${\rm Cov}(\uparrow,\downarrow)=-0.03(1)$
where correlations disappear.

In order to probe the coherence and correlations in momentum space, we further expand the atoms in the horizontal plane
and take the fluorescence image from the microscope. During the expansion, the $z$ lattice is not switched off to keep the atoms at the focus of the objective.
After free expansion in the $y$ direction, the spatial density distribution $\braket{\hat{n}(y)}$ of the atoms is directly related to the momentum distribution $\braket{\hat{n}(q)}$ of the initial state \cite{Bloch2008}. To obtain the distribution of the $\ket{\downarrow}$ component, we push out the $\ket{\uparrow}$ atoms after the TOF. We observe Lorentzian distributions for both atoms and the $\ket{\downarrow}$ atoms in the CSF phase (Fig.~\ref{fig:tof} B) after a 3.5 ms TOF. 

In both of the spin-Mott and CSF phase, the density distributions after TOF show broad peaks which is attributed to the absence of single-particle superfluidity \cite{Kuklov2004,Hu2009}.
Nevertheless, the antipair correlations between the two components could be unveiled from the density-density correlations in the atomic cloud after TOF expansion \cite{Folling2005,Simon2011,Yang2020a,Rosenberg2022,Zheng2022b}. In experiment, we introduce the normalized noise-correlation 
\begin{eqnarray}
    \label{equ:normCor}
   g(d)=\frac{\int\braket{\hat{n}(y)\hat{n}(y+d)}dy} {\int\braket{\hat{n}(y)}\braket{\hat{n}(y+d)} dy}-1,\\
    g_{\sigma\sigma^{\prime}}(d)=\frac{\int\braket{\hat{n}_{\sigma}(y)\hat{n}_{\sigma^{\prime}}(y+d)}dy}{\int\braket{n(y)}\braket{n(y+d)} dy}-\frac{1}{4} \; ,
\end{eqnarray}
where $d$ is the distance. 
Thus, the correlations $g(d)$ and $g_{\downarrow\downarrow}(d)$ 
can be directly derived from the TOF images with both components and single $\ket{\downarrow}$ component, respectively. The inter-component correlation $g_{\uparrow\downarrow}(d)$ is then calculated from the relation $g_{\uparrow\downarrow}(d)=[g(d)-2g_{\downarrow\downarrow}(d)]/2+1/4$.

As shown in Fig.~\ref{fig:tof} C-D, we show the measured noise correlation of $g_{\uparrow\downarrow}(d)$. In the spin-Mott phase, there is no interference peaks at $\pm 2\pi/a_S$. The two components of bosons could be distinguished by their internal states so the Hanbury Brown-Twiss (HBT) effect \cite{Brown1956} breaks down.
In the CSF phase, however, two significant additional peaks at $\pm 2\pi/a_S$ emerge. These peaks do not come from the bosonic quantum statistics but from the antipair correlations $\braket{b_{i,\uparrow}^{\dagger}b_{i,\downarrow}b_{j,\uparrow}b_{j,\downarrow}^{\dagger}}$ and witness the CSF \cite{Hu2009}. 
This unique feature manifests the CSF phase itself from the spin-Mott phase.\\

\begin{figure}
\includegraphics[width=0.99\linewidth]{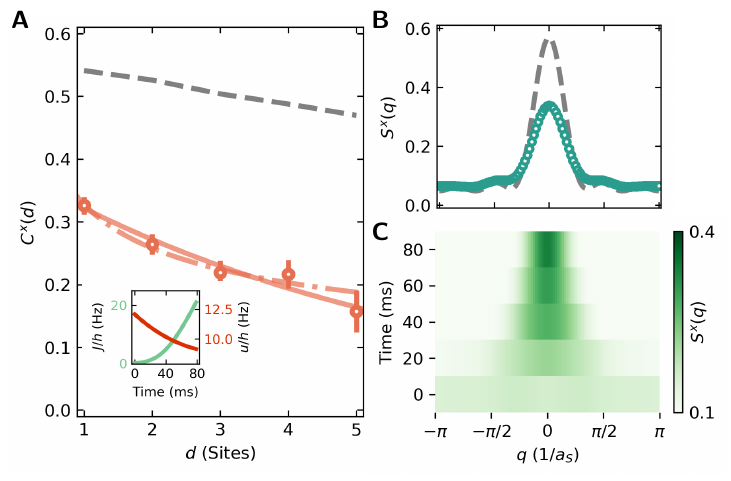}%
\caption{\label{fig:cor}\textbf{Long-range off-diagonal spin correlations.} (\textbf{A}) Long-range off-diagonal spin correlations in the xy-ferromagnetic phase. The correlation length $\xi$ is extracted from the experimental data with an exponential decay fitting (solid orange line). In the $L=6$ chain the correlation length is $\xi=$ 5.9(8) sites, which approaches the system size. For comparison, we also plot an algebra-decay curve fitting (dash-dotted orange line), yielding a Luttinger parameter of 0.7(1). The gray dashed line indicates the predicted correlations of the ground state in the xy-ferromagnetic phase. Inset shows the ramping curve of $J$ (green) and $u$ (red) during the adiabatic passage. (\textbf{B})-(\textbf{C}) Spin structure factors during the phase transition are shown. A peak emerges at the center of the Brillouin zone, which signals the existence of the ferromagnetic order in the $xy$ plane. The gray dashed line in \textbf{B} is the prediction of the ground state in the xy-ferromagnetic phase.
}
\end{figure}

\noindent \textbf{Probing the long-range off-diagonal spin correlations.}
In the following, we further explore the antipair correlations in the CSF 
from the angle of the spin-1 Heisenberg model. 
The off-diagonal spin correlations in the $xy$ plane is defined as
\begin{equation}
\label{equ:offcor}
 \begin{split}
 C^x(d) &=  \braket{ \hat{S}^x_{i} \hat{S}^x_{i+d}}- \braket{\hat{S}^x_{i}}\braket{\hat{S}^x_{i+d}},
\end{split}
\end{equation}
where $d$ is the distance between two lattice sites. 
To detect this off-diagonal spin correlations, we rotate the measurement basis to the $xy$ plane via applying a microwave $\pi/2$ pulse before detection (See SM). This method was widely used in entanglement experiments with atomic qubits \cite{Mandel2003,Dai2016,Yang2020,Zhang2022}. 
Here we adapt it to the spin-1 case (See SM). 
In the spin-Mott state there is no long-range spin correlations,
whereas in the xy-ferromagnetic state 
quasi-long-range spin correlations are expected. 
Fig.~\ref{fig:cor} plots $C^x(d)$ versus distance $d$. 
Using an exponentially decaying ansatz $C^x(d) \propto \exp{(-d/\xi)}$,
we obtain the correlation length $\xi =$ 5.9(8) sites in 
the xy-ferromagnet, comparable to the size of the chain $L=6$. 
This suggests that the temperature of the finite-size system,
achieved in our experiment, is indeed sufficiently low 
for the establishment of a long-range correlation length. 
By comparing the $C^x(1)$ value with that of the finite temperature simulation, we infer that the temperature of the xy-ferromagnet is around $0.5t/k_B$, corresponding to about 1.2 nK.
As expected, the quasi-long-range correlation $C^x(d)$ 
can also be well described by an algebraically decaying function 
(see the dashed line in Fig.~\ref{fig:cor}).
Finally, it is noted that the off-diagonal spin correlations $C^x(d)$ 
of the xy-ferromagnet are related to the antipair correlations $\braket{b_{i,\uparrow}^{\dagger}b_{i,\downarrow}b_{i+d,\uparrow}
b_{i+d,\downarrow}^{\dagger}}$ in the CSF (see SM).

The spin structure factor is an important probe of magnetic materials in condensed matter physics 
\begin{equation}
\label{equ:spinstr}
 \begin{split}
 \mathcal{S}^x(q) &=  \frac{1}{\mathcal{N}}\sum_{i,j}\braket{ \hat{S}^x_{i} \hat{S}^x_{j}}\exp{\left[-\mathrm{i}q\left(i-j\right)a_S\right]},
\end{split}
\end{equation}
where $q\in [-\pi/a_S, \pi/a_S]$ is the wave number. In our experiment, it can be extracted from the real space detection.
As shown in Fig. \ref{fig:cor}, 
a peak around wave number of $q=0$ emerges,
unveiling the buildup of the xy-ferromagnetism \cite{Boothroyd2020}.\\

\noindent \textbf{Conclusion and outlook.}
In conclusion, we observed the CSF in a two-component bosonic mixture in optical lattices. We developed novel procedures to prepare ultra-low entropy spin-Mott state and connected it the the CSF phase adiabatically. Antipair correlations, the hallmark of the CSF, were corroborated by the measurements in both real and momentum spaces under a quantum gas microscope. Furthermore, in the perspective of a spin-1 Heisenberg model, the long-range off-diagonal correlations were extracted by rotating the measurement frame via a MW pulse. The yielded correlation length is 5.9(8) sites and approaches the system size, indicating a low temperature of 0.5$t/k_B$. In the future, our spin-dependent superlattices can be used to study the Su-Schrieffer-Heeger type topological phases \cite{Zheng2017}. We could also tune the spin interaction to be anti-ferromagnetic \cite{Sun2021}, which supports the Haldane phase \cite{Haldane1983,Yang2023}. Applying a magnetic gradient to the CSF will induce coherent spin currents and observation of vortices in a two-dimensional system is possible \cite{Kuklov2004}, which may provide insights to understanding the spin Hall effect.
\\


\let\oldaddcontentsline\addcontentsline
\renewcommand{\addcontentsline}[3]{}


\let\addcontentsline\oldaddcontentsline

\smallskip
\textbf{Acknowledgements}
We acknowledged Long-Xiang Liu and Nikolay Prokof'ev for fruitful discussions. This work was supported by the National Natural Science Foundation of China (Grant No. 12125409), the  Innovation Program for Quantum Science and Technology (2021ZD0302004), and the Anhui Initiative in Quantum Information Technologies. YGZ acknowledged the support by the Fundamental Research Funds for the Central Universities, the CPS-Huawei MindSpore Fellowship, and the China Postdoctoral Science Foundation (2023TQ0102).

\setcounter{figure}{0}
\renewcommand{\thefigure}{S\arabic{figure}}
\setcounter{equation}{0}
\renewcommand{\theequation}{S\arabic{equation}}
\setcounter{table}{0}
\renewcommand{\thetable}{S\arabic{table}}

\renewcommand{\thesection}{\arabic{section}}
\clearpage

\onecolumngrid

\begin{center}
	\textbf{\large SUPPLEMENTARY MATERIALS}
\end{center}
\normalsize


\tableofcontents
\newpage
\twocolumngrid

\section{Experimental sequences}

\begin{figure*}[htbp] 
	\centering
	\includegraphics[width=1.\textwidth]{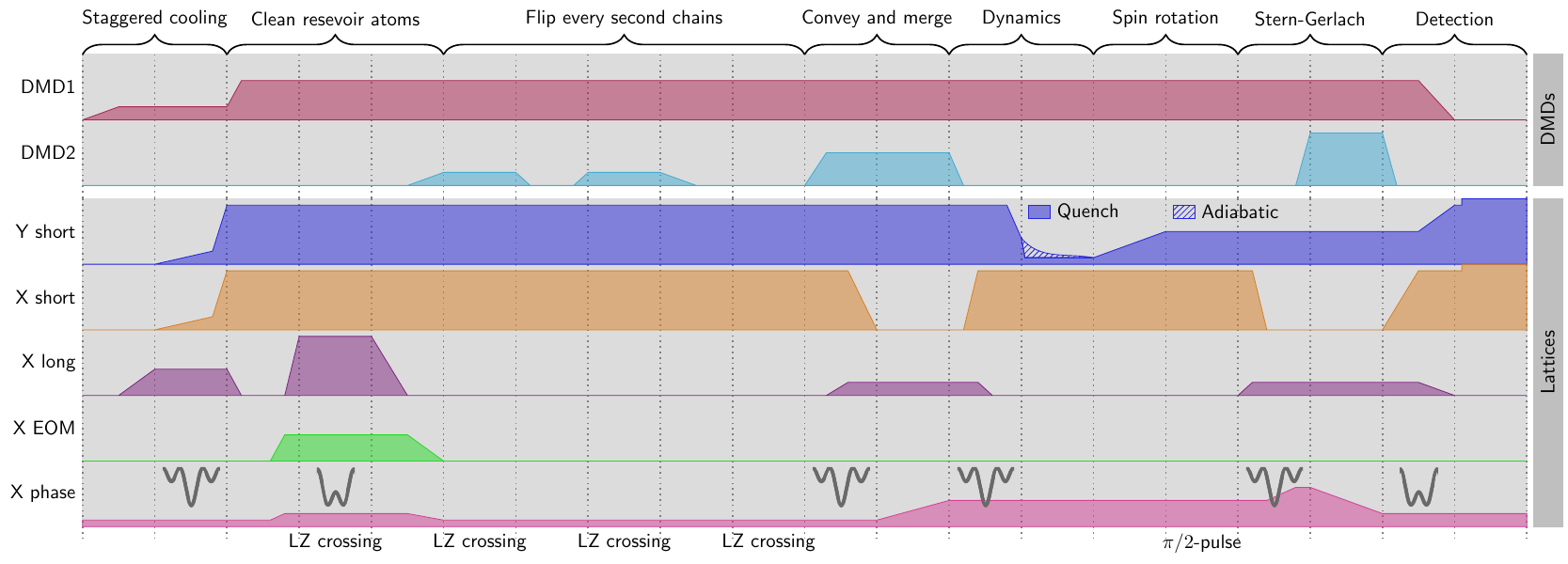}
	\caption{\textbf{Experimental sequences.} Some most relevant signals for the experiments are shown. Note that the time durations of the steps are not proportional to their lengths. The DMD1 is used for cancellation of the harmonic trapping potential induced by red-detuned lattice beams. The DMD2 is used for projection of addressing beams. X EOM refers to the voltage signal driving the electro-optic modulator to rotate the polarization of the $x$ long lattice. X phase denotes the relative phase control of the $x$ long lattice. We utlized the Landau-Zener crossing for high fidelity spin flip. For measurements of off-diagonal spin correlations, a spin rotation ($\pi/2$-pulse) is inserted before the Stern-Gerlach separation.}
	\label{fig:seq}
\end{figure*}

\subsection{Preparation of doublon occupied chains}
Fig. \ref{fig:seq} shows the brief sketch of the experimental sequence for part of the signals. After loading the Bose-Einstein condensate (BEC) of ${}^{87}$Rb atoms to the $z$ lattice, we selected a single layer of the atoms \cite{Xiao2020}. In the two-dimensional system, we further evaporatively cool the atoms to BEC. Before ramping up the $xy$ short lattices to drive the transition to a Mott insulator, we impose a long lattice in the $x$ direction for staggered immersion cooling during the transition \cite{Yang2020}. The atoms in the reservoir are pushed out after being addressed and flipped to the $\ket{\uparrow}$ state assisted by the spin-dependent superlattice in the $x$ direction. Next, we convey every second chains for two sites to merge with the stayed chains. This moving is realized by varying the phase of the $x$ long lattice to be a conveyor belt for the atoms \cite{Li2021}. The atoms in stayed chains are pinned by another addressing beam projected from a DMD during the conveying. Before the conveying, the stayed atoms are flipped to $\ket{\uparrow}$ state. We choose a tune-out wavelength for the addressing beam which has no Stark shift for the $\ket{\downarrow}$ atoms \cite{Zheng2022}. Finally, we prepared around six doublon occupied chains, where each site are occupied by a pair of $\ket{\uparrow}$ and $\ket{\uparrow}$ atoms. Please note that, now the doublon occupied chains are separated by four short lattice sites. These unoccupied sites are used for later spin- and number-resolved detection. During the conveying we switched off the $x$ short lattice and after the merge we loaded the atom back to the $x$ short lattice for the following physics experiments.

\subsection{Spin-resolved readout of the doublon occupied chains}
After adiabatic passage or quenched evolution, we freeze the atoms for detection. Conventional fluorescence imaging ignores the information of the atoms' internal states \cite{Bakr2009}. To resolve the internal states, we need to split the atoms with opposite spins spatially, that is, Stern-Gerlach detection, similar to this work \cite{Boll2016} and our previous work \cite{Zhang2022}. In our case, this separation is realized by conveying of $\ket{\downarrow}$ atoms while pinning the $\ket{\uparrow}$ atoms by the addressing beam, which is similar to the conveying process in the preparation of the initial states.

However, there are still some sites occupied with two identical spins in the doublon occupation case and those sites will undergo pair-wise loss during the fluorescence imaging. So another spin-independent separation is performed to remove possible doublons before the fluorescence imaging. After we convey the atoms for a distance of 2.5 short lattice sites to separate the two spins, we further ramp up the $x$ short lattice to split the two atoms in the balanced double wells. Now the two particles are distributed across four lattice sites in the $x$ direction. From the spatial distribution of the two atoms in four sites, we are able to deduce the outcome in every site of the doublon occupied chain.

\section{Detection of the off-diagonal spin correlations}

\subsection{Off-diagonal spin correlations}
To detect the xy-ferromagnetic state we need to probe off-diagonal correlation function $\langle \hat{S}_i^{+}  \hat{S}_{i+l}^- \rangle$ \cite{Schachenmayer2015}. Since $\hat{S}^{\pm}=\hat{S}^x\pm \mathrm{i}\hat{S}^y$
\begin{equation}
\label{equ:spsm}
\begin{split}
  \hat{S}_i^{+}  \hat{S}_{i+l}^- =& \hat{S}_i^x \hat{S}_{i+l}^x +\mathrm{i}\hat{S}_i^y \hat{S}_{i+l}^x\\
  &-\mathrm{i}\hat{S}_i^x \hat{S}_{i+l}^y+\hat{S}_i^y \hat{S}_{i+l}^y.
  \end{split}
\end{equation}
According to the commuting relations $[\hat{S}_i^x,\hat{S}_j^y]=\mathrm{i}\hat{S}_i^z\delta_{ij}$, when $l\neq0$
\begin{equation}
\label{equ:spsm1}
\begin{split}
    \langle \hat{S}_i^+ \hat{S}_{i+l}^- \rangle=&\langle \hat{S}_i^x \hat{S}_{i+l}^x \rangle +\langle \hat{S}_i^y \hat{S}_{i+l}^y \rangle \\
    &+\mathrm{i}\langle \hat{S}_{i+l}^x \hat{S}_i^y \rangle -\mathrm{i}\langle \hat{S}_i^x \hat{S}_{i+l}^y \rangle\\
  =&  \langle \hat{S}_i^{x}  \hat{S}_{i+l}^x \rangle+  \langle \hat{S}_i^{y}  \hat{S}_{i+l}^y\rangle.
\end{split}
\end{equation}
The cross terms are cancelled owing to the translation symmetry and space inversion symmetry. If $l=0$, 
\begin{equation}
\label{equ:spsm2}
  \langle \hat{S}_i^+ \hat{S}_{i}^- \rangle=\langle (\hat{S}_i^x)^2\rangle + \langle (\hat{S}_i^y)^2 \rangle + \langle \hat{S}_i^z \rangle.
\end{equation}
Combining Eq. \ref{equ:spsm1} with Eq. \ref{equ:spsm2} we obtain that 
\begin{equation}
\label{equ:spsm3}
  \langle \hat{S}_i^+ \hat{S}_{j}^- \rangle=\langle \hat{S}_i^x \hat{S}_j^x\rangle + \langle \hat{S}_i^y \hat{S}_y^y \rangle,
\end{equation}
where $\langle \hat{S}_i^z\rangle=0$ for $S^z=0$ sector. Due to the $U(1)$ symmetry in the $xy$ plane, $\langle \hat{S}_i^x \hat{S}_j^x\rangle =\langle \hat{S}_i^y \hat{S}_y^y \rangle$, thus $\langle \hat{S}_i^+ \hat{S}_{j}^- \rangle=2\langle \hat{S}_i^x \hat{S}_j^x\rangle=2\langle \hat{S}_i^y \hat{S}_j^y\rangle$.

\subsection{Implementation in experiments}
In experiment the measurements is performed in $ \hat{S}^z $ basis. To obtain the off-diagonal expectation values, 
it is equivalent to diagonal measurements after applying a unitary operator on the states
\begin{equation}
  \begin{split}
    \langle \phi_i \mid \hat{S}^y \mid \phi_i \rangle &= \langle \phi_i \mid \hat{U}^{\dagger} \hat{S}^z \hat{U} \mid \phi_i \rangle\\
    &=\langle \phi_f \mid \hat{S}^z \mid \phi_f \rangle .
  \end{split}
  \label{equ:OffdiagonalSy}
\end{equation}
Now the question is to find such an operator which satisfies 
\begin{equation}
  \hat{U} \hat{S}^y \hat{U}^{\dagger} =\hat{S}^z.
  \label{equ:USyUSz}
\end{equation}
Since $\hat{S}^y$ and $\hat{S}^z$ have the same eigenvalues, Eq. \ref{equ:USyUSz} is diagonalization of the $\hat{S}^y$ to $\hat{S}^z$ if there exists $\tilde{\hat{U}}=\hat{U}^{\dagger},\tilde{\hat{U}}^{-1}=\hat{U}$. To diagonalize $\hat{S}^y$, $\tilde{\hat{U}}$ is formed with eigenvectors
\begin{equation}
  \tilde{\hat{U}}=\frac{1}{2}
  \begin{pmatrix}
    1 & \sqrt{2} \mathrm{i} & -1 \\
    \sqrt{2} \mathrm{i} & 0 & \sqrt{2} \mathrm{i}\\
    -1 & \sqrt{2} \mathrm{i} & 1
  \end{pmatrix}.
\end{equation}
And the inverse matrix 
\begin{equation}
  \tilde{\hat{U}}^{-1}=\frac{1}{2}
  \begin{pmatrix}
    1 & -\sqrt{2} \mathrm{i} & -1 \\
    -\sqrt{2} \mathrm{i} & 0 & -\sqrt{2} \mathrm{i}\\
    -1 & -\sqrt{2} \mathrm{i} & 1
  \end{pmatrix}
  =\tilde{\hat{U}}^{\dagger}.
\end{equation}
The spin-1 matrices are 
\begin{equation}
\begin{split}
  &\hat{S}^x=\frac{1}{\sqrt{2}}
  \begin{pmatrix}
    0 & 1 & 0 \\
    1 & 0 & 1 \\
    0 & 1 & 0
  \end{pmatrix},\\
  &\hat{S}^y=\frac{1}{\sqrt{2}}
  \begin{pmatrix}
    0 & -\mathrm{i} & 0 \\
    \mathrm{i} & 0 & -\mathrm{i}\\
    0 & \mathrm{i} & 0
  \end{pmatrix},\\
  &\hat{S}^z=
  \begin{pmatrix}
    1 & 0 & 0 \\
    0 & 0 & 0 \\
    0 & 0 & -1
  \end{pmatrix}.
\end{split}
\end{equation}
It is easy to verify that
\begin{equation}
  \tilde{\hat{U}}^{-1} \hat{S}^y \tilde{\hat{U}}=\hat{S}^z.
\end{equation}
Thus $\hat{U}=\tilde{\hat{U}}^{-1}$ is obtained and $\hat{S}^y=\hat{U}^{\dagger}\hat{S}^z \hat{U}$.
In the following section, we can find that applying a $\pi/2$ microwave pulse is $\hat{U}$. In our experiments, two spin-$\frac{1}{2}$ particles constitute one spin-1,
\begin{equation}
\begin{split}
  \ket{ 1 }=&\ket{ \frac{1}{2} ,\frac{1}{2} },\\
  \ket{ 0}=&\frac{1}{\sqrt{2}}\left( \ket{ \frac{1}{2} ,-\frac{1}{2} } + \ket{ -\frac{1}{2} ,\frac{1}{2} } \right),\\
  \ket{ -1 }=&\ket{ -\frac{1}{2} ,-\frac{1}{2} }.
  \end{split}
\end{equation}
Applying a $\pi/2$ microwave pulse to the state $\ket{ 1 }$

\begin{equation}
  \hat{U}_1 \otimes \hat{U}_2 \ket{ 1 } = \hat{U}_1\otimes \hat{U}_2 \ket{ \frac{1}{2} ,\frac{1}{2} },
\end{equation}
where $\hat{U}_1=\hat{U}_2=e^{-\mathrm{i}\pi \hat{\sigma}^x /4}$ is a $\pi/2$ microwave pulse coupling the two levels of a spin-$\frac{1}{2}$. Its matrix form is

\begin{equation}
  \hat{U}_1=e^{-\mathrm{i}\pi \hat{\sigma}^x /4}=\frac{1}{\sqrt{2}}
  \begin{pmatrix}
    1 & -\mathrm{i}\\
    -\mathrm{i} & 1
  \end{pmatrix}.
\end{equation}
And the states 
\begin{equation}
  \ket{ \frac{1}{2}}=
  \begin{pmatrix}
    1\\
    0
  \end{pmatrix},
  \ket{ -\frac{1}{2} }=
  \begin{pmatrix}
    0\\
    1
  \end{pmatrix}.
\end{equation}
Now the state becomes
\begin{equation}
\begin{split}
  \hat{U}_1 \otimes \hat{U}_2 \ket{ 1 }=&\frac{1}{2}\left[ \ket{ \frac{1}{2},\frac{1}{2}} +\mathrm{i}\left( \ket{ \frac{1}{2} ,-\frac{1}{2} }  + \ket{ -\frac{1}{2} ,\frac{1}{2} } \right)- \ket{ -\frac{1}{2} ,-\frac{1}{2} } \right]\\
  =&\frac{1}{2}\left( \ket{ 1 } -\mathrm{i}\sqrt{2}\ket{ 0 }-\ket{ -1 }\right)
\end{split}.
  \label{equ:U1U21}
\end{equation}
On the other hand, if we apply $\hat{U}$ to the spin-1 state
\begin{equation}
  \hat{U}\ket{ 1}= \frac{1}{2}\left( \ket{ 1 } -\mathrm{i}\sqrt{2}\ket{ 0 }-\ket{ -1 }\right),
\end{equation}
which is indentical to what we got in \ref{equ:U1U21}. So do the other two states.
\begin{equation}
\begin{split}
  \hat{U}\ket{ 0}=\hat{U}_1 \otimes \hat{U}_2 \ket{ 0 } =& -\frac{\mathrm{i}}{\sqrt{2}}\left( \ket{ 1} + \ket{ -1 } \right),\\
  \hat{U}\ket{ -1}=\hat{U}_1 \otimes \hat{U}_2 \ket{ -1}=&-\frac{1}{2}\left( \ket{ 1 } + \mathrm{i}\sqrt{2}\ket{ 0} -\ket{ -1 }\right).
  \end{split}
\end{equation}

\begin{figure}[tbp!] 
	\centering
	\includegraphics[width=0.49\textwidth]{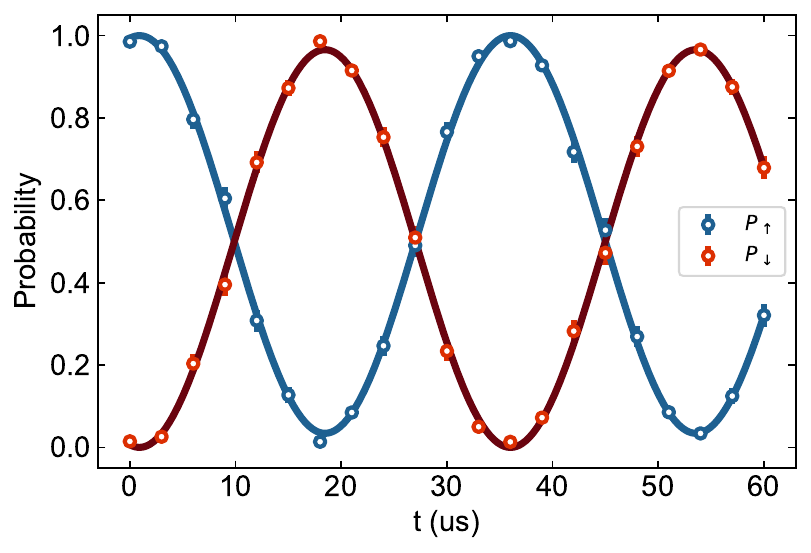}
	\caption{Microwave driving Rabi oscillation. The atoms are initialised in $\ket{\uparrow}$ state and we applied a resonant microwave pulse to drive the oscillation of the populations between $\ket{\uparrow}$ (blue circles) and $\ket{\downarrow}$ (red circles) states. The solid lines are fitting with sinusoidal curves.}
	\label{fig:Rabi}
\end{figure}

From the point view of spin rotation, $\hat{U}$ is a spin rotation along the $\hat{S}^x$ axis
\begin{equation}
  \hat{U}=e^{-\mathrm{i}\theta \hat{S}^x},
\end{equation}
$\hat{S}^z$ under the rotation is 
\begin{equation}
\begin{split}
  \hat{U}^{\dagger}\hat{S}^z \hat{U}=&e^{\mathrm{i}\theta \hat{S}^x} \hat{S}^z  e^{-\mathrm{i}\theta \hat{S}^x} \\
  =&\hat{S}^z +\mathrm{i}\theta [\hat{S}^x,\hat{S}^z] +\frac{1}{2!}(\mathrm{i}\theta)^2[\hat{S}^x,[\hat{S}^x,\hat{S}^z]]+\cdots\\
  =&\hat{S}^z \cos \theta +\hat{S}^y \sin \theta   ,
\end{split}
\end{equation}
If $\theta=\pi/2$, $\hat{U}=e^{-\mathrm{i}\pi \hat{S}^x/2}$. For spin-1 case, 
\begin{equation}
  e^{\mathrm{i}\theta \hat{S}^x}=I+\mathrm{i}\hat{S}^x \sin \theta +(\hat{S}^x)^2\cos \theta -(\hat{S}^x)^2.
\end{equation}
So we have 
\begin{equation}
\begin{split}
  e^{-\mathrm{i}\pi \hat{S}^x/2}=&I-\mathrm{i}\hat{S}^x -(\hat{S}^x)^2\\
  =& \frac{1}{2}
  \begin{pmatrix}
    1 & -\sqrt{2} \mathrm{i} & -1 \\
    -\sqrt{2} \mathrm{i} & 0 & -\sqrt{2} \mathrm{i}\\
    -1 & -\sqrt{2} \mathrm{i} & 1
  \end{pmatrix}\\
  =&\hat{U}.
  \end{split}
\end{equation}
Besides, the above formula is also valid for spin-$\frac{1}{2}$ case, $\hat{S}^x=\hat{\sigma}^x/2$,
\begin{equation}
\begin{split}
  e^{-\mathrm{i}\pi \hat{S}^x/2}=&e^{-\mathrm{i}\pi \hat{\sigma}^x/4}\\
  =& \frac{1}{\sqrt{2}}\left(I -\mathrm{i}\hat{\sigma}^x \right)\\
  =& \hat{U}_1.
  \end{split}
\end{equation}

\subsection{Microwave pulse}
In the experiments, we applied an $\pi/2$-pulse to rotate the spin to the $xy$ plane. The length of the $\pi/2$-pulse is determined from the Rabi oscillation between the two components driving by the microwave. The Rabi frequency is $2\pi \times$26 kHz, corresponding to a $\pi/2$-pulse of 9.6 $\mu$s. During the oscillation, the magnetic field is stablized to achieve a high fidelity for the spin rotation.

\section{Time-of-flight measurements}
\subsection{Momentum distribution and noise-correlations}
The momentum distribution of the two-component Bose-Hubbard chain is 
\begin{equation}
\label{equ:ndtof}
    \braket{\hat{n}_{\sigma}(q)}=\frac{1}{\mathcal{N}}\sum_{i,j}\braket{b_{i,\sigma}^{\dagger}b_{j,\sigma}}e^{\mathrm{i}q(i-j)a_S},
\end{equation}
where $\mathcal{N}$ is the normalized factor. The total density distribution of the two-components is the sum of each component $\braket{\hat{n}(q)}=\braket{\hat{n}_{\uparrow}(q)}+\braket{\hat{n}_{\downarrow}(q)}$. In the spin-Mott phase, there is no long-range correlations $\braket{b_{i,\sigma}^{\dagger}b_{j,\sigma}}=\delta_{i,j}$. Therefore, the momentum distribution is flat. On the contrary, the existence of the antipair correlations correlation will modify the line shape of the distribution, where a peak emerges in the center located at $q=0$.

\begin{figure*}[tbp!] 
	\centering
	\includegraphics[width=0.8\textwidth]{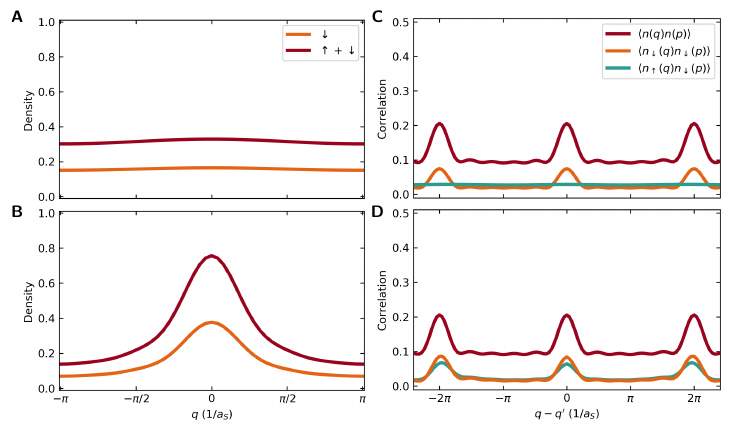}
	\caption{Momentum distribution and correlations in the spin-Mott and CSF phases. We calculate the density distribution and correlations from the ground states of the two phases obtained by DMRG numerical simulations. A and B are the momentum distribution in the spin-Mott and CSF phases. C and D show the noise correlations in the spin-Mott and CSF phases, respectively.}
	\label{fig:Momdis}
\end{figure*}

The density correlation in momentum space is 
\begin{equation}
\label{equ:nntof}
\begin{split}
    \braket{\hat{n}_{\sigma}(q_1)\hat{n}_{\sigma^{\prime}}(q_2)}=\frac{1}{\mathcal{N}^2}\sum_{i,j,l,m}\braket{b_{i,\sigma}^{\dagger}b_{j,\sigma}b_{l,\sigma^{\prime}}^{\dagger}b_{m,\sigma^{\prime}}}\\
    \times e^{\mathrm{i}q_1(i-j)a_S}e^{\mathrm{i}q_2(l-m)a_S}.
\end{split}
\end{equation}
In the spin-Mott phase, owing to the periodic structure of the lattice, there are peaks located at $2M\pi/a_S$ in the intra-component correlations $\braket{\hat{n}_{\uparrow}(q_1)\hat{n}_{\uparrow}(q_2)}$ and $\braket{\hat{n}_{\downarrow}(q_1)\hat{n}_{\downarrow}(q_2)}$, where $M$ is an arbitrary integer number. This is the so-called HBT interference of bosons. However, there is no peaks in the correlation of inter-component atoms $\braket{\hat{n}_{\uparrow}(q_1)\hat{n}_{\downarrow}(q_2)}$ and $\braket{\hat{n}_{\downarrow}(q_1)\hat{n}_{\uparrow}(q_2)}$. In this case, the particles are distinguishable by their internal states and thus the HBT interference peaks disappear. The total density correlation is the sum of the intra- and inter-component correlations $\braket{n(q_1)n(q_2)}=2[\braket{\hat{n}_{\downarrow}(q_1)\hat{n}_{\downarrow}(q_2)}+\braket{\hat{n}_{\uparrow}(q_1)\hat{n}_{\downarrow}(q_2)}]$ and shows similar peaks in the $2M\pi/a_S$. In contrast, additional peaks emerge in the inter-component correlations in the CSF phase due to the antipair correlations between the two-component atoms $\braket{b_{i,\uparrow}^{\dagger}b_{i,\downarrow}b_{j,\uparrow}b_{j,\downarrow}^{\dagger}}$. We show the numerical results in Fig. \ref{fig:Momdis}.

\subsection{TOF measurements}
We perform TOF expansion of the atoms by abruptly switching off the $xy$ lattices while keeping the $z$ lattice 40$E_r$ to avoid the atoms out of the focus of the objective in the imaging axis. After 1 ms expansion we switch on the $x$ lattice back to stop the expansion in the $x$ direction and continue the flight in the $y$ direction for another 2.5 ms. The first 1 ms expansion in the $x$ direction reduces the interaction of the doublons and also decreases the probability of doublons during the florescence detection. To probe the density distribution and noise correlation of the $\ket{\downarrow}$ component, we push out the $\ket{\uparrow}$ atoms with a resonant light pulse before the florescence imaging. In the TOF experiment, we address only one chain in each shot, and post-select the realizations with atoms between 10 and 12 for both components images (half of the numbers for single component images). The data in Fig. \ref{fig:tof} are averaged over 200-500 chains. To suppress the noise induced by the atom number fluctuations, we restricted the analysis in a region of interest of 60 sites in the center and binned every 5 sites. The global gaussian profile of the expanded atom cloud is attributed to the Wannier function of the atoms localized in the lattice sites before expansion.

\subsection{Harmonic confinement during the TOF}
As we do not switch off all the vertical lattice during the TOF, the atoms expands in the confinement of the $z$ lattice potentials, inducing a correction to the correspondence between the in trap momentum and the spatial position after a fixed TOF time. The atoms will oscillate in the harmonic trap like a cradle. The amplitude of the oscillation $A_y$ depends on the initial kinetic energy of the atom
\begin{equation}
    \label{equ:tofharm}
    \frac{1}{2}m_{\mathrm{Rb}}\omega_y^2 A_y^2=\frac{\hbar^2 q^2}{2m_{\mathrm{Rb}}},
\end{equation}
where $\omega_y$ is the trap frequency. At the time $t$, the location of the atom after TOF is 
\begin{equation}
    \label{equ:AmpTof}
    A_y\sin{\omega_y t}=\frac{\hbar q}{m_{\mathrm{Rb}}\omega_y}\sin{\omega_y t},
\end{equation}
which is also proportional to the initial momentum $q$. The calibrated harmonic trapping frequency is $2\pi \times$38.0(6) Hz via the breathing mode oscillation \cite{Stringari1996}. For a TOF time of 3.5 ms, the predicted location of the $2\pi/a_S$ peak is $d=$ 36 sites, which agrees well with the experimental data.

\begin{figure}[tbp!] 
	\centering
	\includegraphics[width=0.45\textwidth]{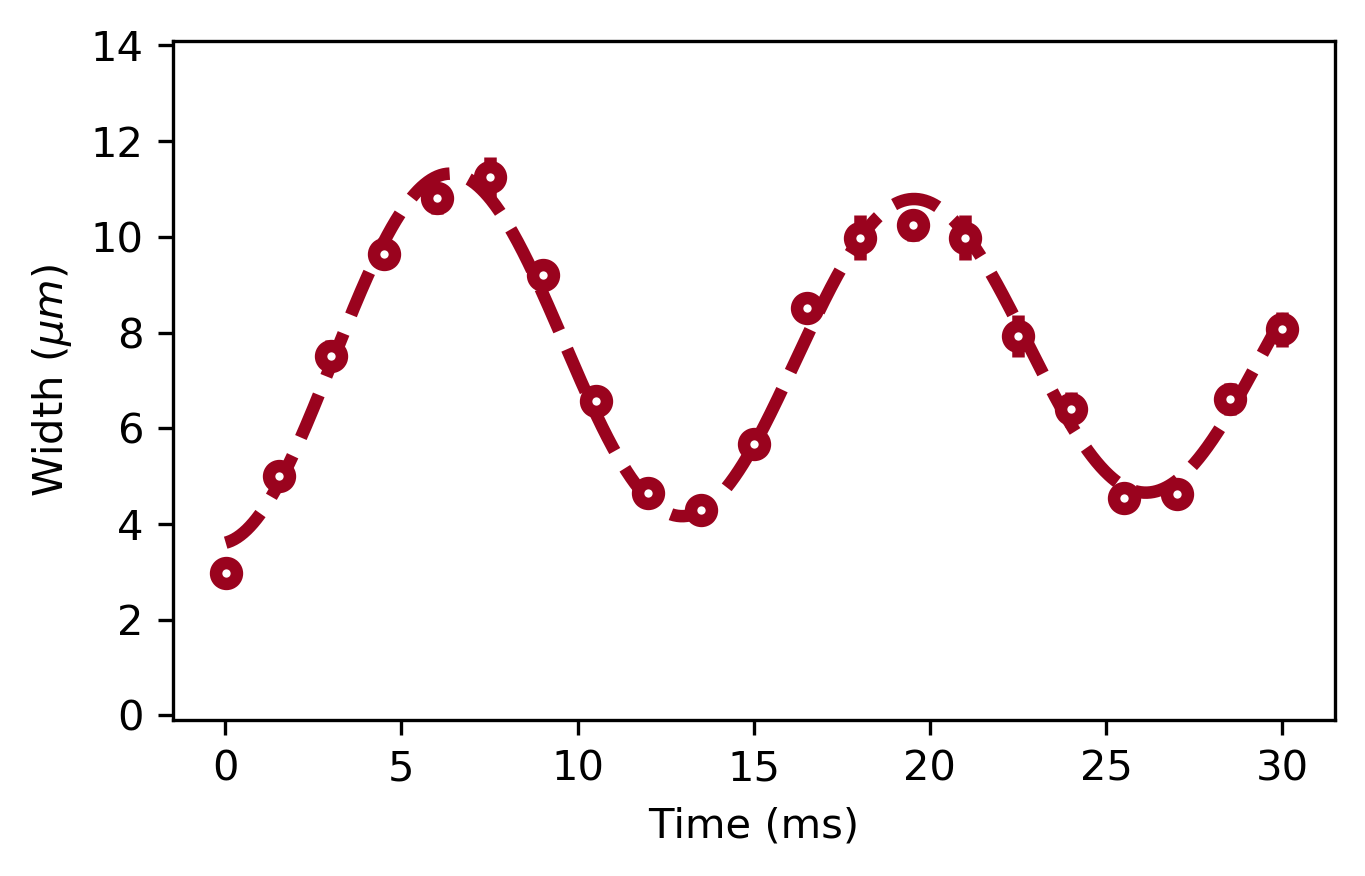}
	\caption{Breathing mode oscillation. We excite the breathing oscillation and monitor the width of the atomic cloud along the $y$ lattice direction at various times. The dashed line is an sinusoidal curve fitting and the oscillation frequency is 76.0(6) Hz, yielding a trap frequency of $\omega_y/2\pi=$ 38.0(3) Hz.}
	\label{fig:Breathing}
\end{figure}

\section{From two-component BHM to Heisenberg model}

The system can be well described by the two-component BHM. To simplify the analysis, we consider a two-site model without loss of generality. The two-site Hamiltonian is given by:
\begin{equation} \label{eq1}
\begin{split}
     H_{BHM} = &-\sum_{\sigma}(t_\sigma b_{1,\sigma}^\dagger b_{2,\sigma} + h.c.) \\& + \sum_{\sigma,\sigma^\prime \atop i=1,2}\frac{U_{\sigma\sigma^\prime}}{2}b_{i,\sigma}^\dagger b_{i,\sigma^\prime}^\dagger b_{i,\sigma^\prime}b_{i,\sigma}\\ &+ \sum_{\sigma\atop i=1,2}\mu_{i,\sigma} \hat{n}_{i,\sigma},
\end{split}
\end{equation}
where $\sigma =\{\uparrow ,\downarrow\}$ denotes the spin of atoms,  $t_\sigma$ represents the tunnelling strength of the $\sigma$-component atoms, $U_{\sigma\sigma^\prime}$ is the onsite interaction strength between atoms of different spin $\sigma$ and $\sigma^\prime$, and $\mu_{i,\sigma}$ is the chemical potential. In the limit where the interaction energy is much larger than the tunnelling strength, i.e., $U \gg t$, the particle tunnelling process is suppressed. In this case the low energy Hilbert space can be described by the state $|g_i\rangle = |n_L^\uparrow, n_L^\downarrow; n_R^\uparrow, n_R^\downarrow \rangle$ with fixed $N_L = n_L^\uparrow+ n_L^\downarrow$, $N_R = n_R^\uparrow+ n_R^\downarrow$. However, coupling of different $|g\rangle$ states can still occur through higher-order processes. To obtain the effective Hamiltonian of the system, we treat the tunnelling term as a perturbation and use the second-order perturbation theory:
\begin{equation} \label{eq2}
    H_{eff} = H_0 + \sum_{i,j,k}\frac{\bra {g_j} V \ket{e_k} \bra{e_k} V \ket{g_i} }{E-\bra{e_k}H_0\ket{e_k}}\ket{g_j}\bra{g_i},
\end{equation}
where $H_0$ is the onsite interaction term in the original Hamiltonian $H_{BHM}$, the the second term is the coupling term between different quantum states due to the second-order perturbation, and $V$ is the tunnelling term in $H_{BHM}$. The states in the high-energy subspace is denoted by $\ket{e_k}$. Substituting Eq. \ref{eq1} into Eq. \ref{eq2}, we obtain the effective Hamiltonian as:
\begin{equation}
\begin{split}
 H_{eff} &= H_0 \\
 & -\sum_{i\neq j \atop \sigma,\sigma^\prime } \frac{t_{\sigma^\prime}t_{\sigma} b_{j,\sigma^\prime}^\dagger b_{i,\sigma^\prime}b_{i,\sigma}^\dagger b_{j,\sigma}}{U_{\sigma\sigma^\prime}(\hat{n}_j^{\sigma^\prime} - \hat{n}_i^{\sigma^\prime}) + U_{\sigma\sigma}(\hat{n}_j^\sigma- \hat{n}_i^\sigma+1)}.
\end{split}
\end{equation}
In our system, the atom number of each site is fixed to $N_L=N_R=2$. We can map the particle number representation to the spin representation via: $\hat{S}^+ = b_\uparrow^\dagger b_\downarrow, \hat{S}^- = b_\downarrow^\dagger b_\uparrow $, $2\hat{S}_z =\hat{n}_\uparrow - \hat{n}_\downarrow $. In this case, the original Bose-Hubbard Hamiltonian is mapped to a spin-1 Heisenberg Hamiltonian
 \begin{equation}
 \label{eq:S1HM1}
     \begin{split}
        H = &-\sum_{i,j} (J_\perp (\hat{S}_i^x\hat{S}_j^x+\hat{S}_i^y\hat{S}_j^y) + J_z \hat{S}_i^z\hat{S}_j^z )\\ & + u\sum_{i}(\hat{S}_i^z)^2 + h\sum_{i}\hat{S}_i^z,
    \end{split}
 \end{equation}
where $u$ is the uniaxial single-ion anisotropic term of the spin-1 model and $h$ is the effective magnetic field. The coupling terms are given by:
\begin{equation}
    \begin{split}
        &J_\perp=\frac{4t_\uparrow^2}{U_{\uparrow \downarrow}}\\
        &J_z=2(\frac{t_\uparrow^2}{U_{\uparrow \uparrow}} +\frac{t_\downarrow^2}{U_{\downarrow \downarrow}}) \\
        &u=\frac{U_{\uparrow \uparrow}+U_{\downarrow \downarrow}-2U_{\uparrow \downarrow}}{2}\\
        &h =(\mu_{i,\uparrow}-\mu_{i,\downarrow})
    \end{split}
\end{equation}

For the two components considered here, the two intra-component interactions are nearly identical (differ less than $\sim 0.1\%$) \cite{Stamper-Kurn2013a}, so we assume $U_{\uparrow \uparrow}=U_{\downarrow \downarrow}=U$. The anisotropy is also small $u=U-U_{\uparrow\downarrow}\sim~0.01U$. Besides, the trap depths of the optical lattices for the two components are identical as well, yielding a zero effective magnetic field $h=0$. And the tunnelling strengths are also identical for the two components, i.e., $t_{\uparrow}=t_{\downarrow}=t$. Now the spin-1 Heisenberg model Eq. \ref{eq:S1HM1} is simplified to 

\begin{equation}
\label{equ:S1HM}
 \begin{split}
 \mathcal{\hat{H}} &=  -J \sum_{i} \mathbf{S}_{i} \cdot \mathbf{S}_{i+1}+ u \sum_{i} \left(\hat{S}_{i}^z\right)^2 \; ,
\end{split}
\end{equation} 
where $J=J_\perp=J_z=4t^2/U_{\uparrow\downarrow}$. When taking the corrections from the non-standard BHM \cite{Dutta2015} into account, the superexchange coupling is enhanced to $J=4[t+(\hat{n}_i+\hat{n}_j-1)T]^2/U_{\uparrow\downarrow}$, where $T$ denotes the density induced tunnelling. For spin-1 case with two atoms per site, it is $J=4(t+3T)^2/U_{\uparrow\downarrow}$, whereas for spin-1/2 case with one atom per site, it is $J=4(t+T)^2/U_{\uparrow\downarrow}$.
\begin{figure}[tbp!] 
	\centering
	\includegraphics[width=0.45\textwidth]{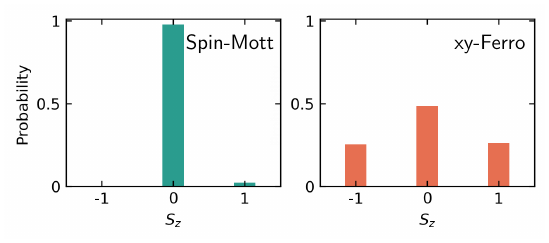}
	\caption{Full-count statistics of the local magnetization in the spin-Mott (left panel) and xy-ferromagnet phases.}
	\label{fig:CntSz}
\end{figure}

\begin{figure}[tbp!] 
	\centering
	\includegraphics[width=0.45\textwidth]{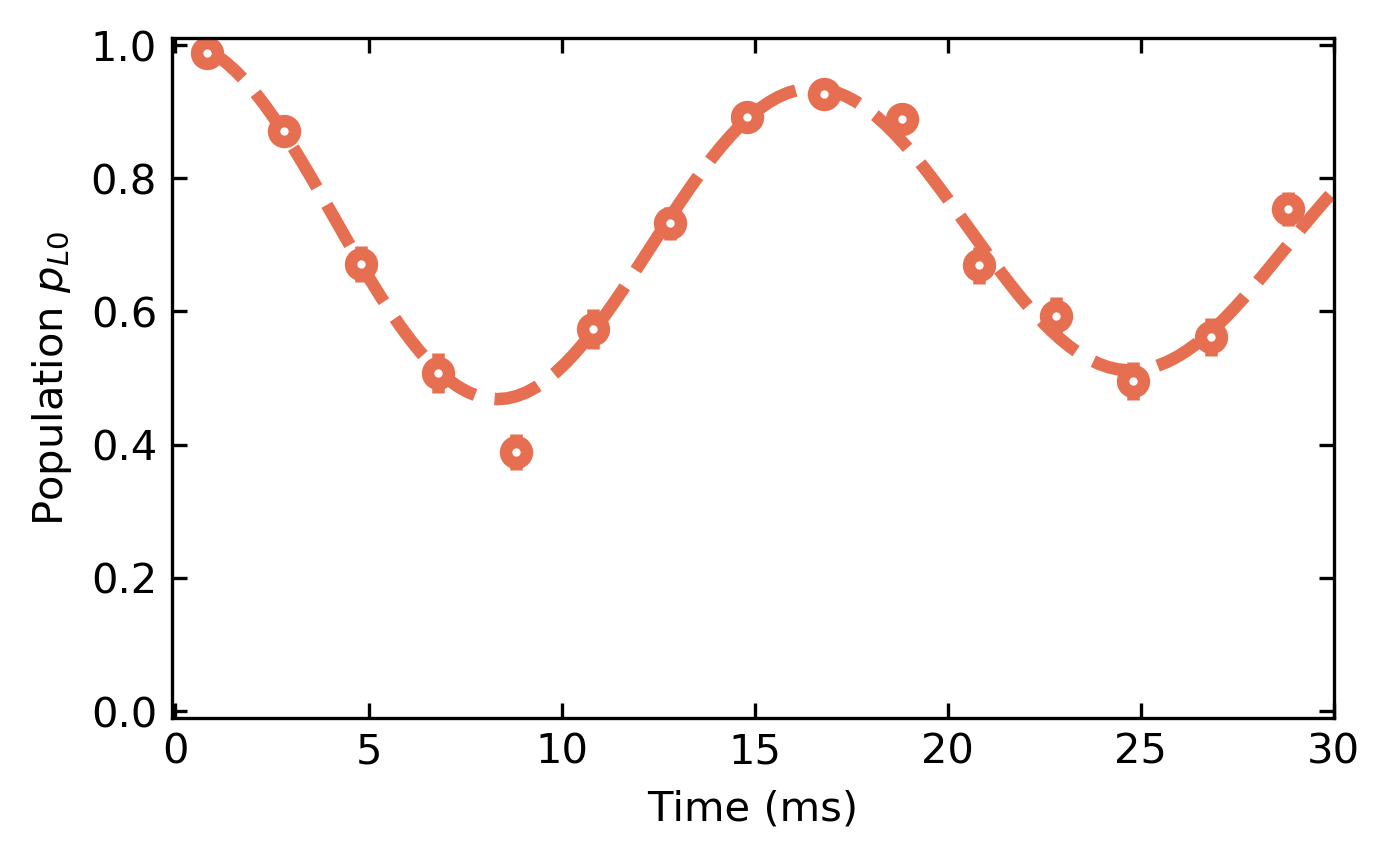}
	\caption{Four-particle dynamics in double wells. We prepared the atoms in to $\ket{S_z=0}$ state in an array of double wells by imposing the $y$ long lattice. Then the $y$ short lattice is quenched to 10$E_r$ to start the dynamics in double wells. The dashed line is a sinusoidal curve fitting from which we inferred the anisotropy $u$.}
	\label{fig:DWdyn}
\end{figure}

\section{Scattering lengths of inter- and intra-component atoms}

We calibrated the single-ion term $u$ in experiments. We initialized the atoms to $\ket{S_z=0}$ state in double wells and observed the population evolution $p_{L0}$, where $p_{L0}$ is the population of $\ket{S_z=0}$ in the left wells. In such a simple case of two-site model, we could directly derive the analytical form of the dynamics,
\begin{equation}
\begin{split}
    p_{L0}(t)=&2A(1-A)\cos[2w(t-t_0)]e^{-\frac{t}{\tau}}+\\
    &(1-A)^2 e^{-\frac{2t}{\tau}}+A^2,
\end{split}
    \label{equ:Dwdyn}
\end{equation}
where $A=(\Delta+w)/2w$, $w=\sqrt{\Delta^2+2J^2}$ and $\Delta=u+J/2$. We also took an exponential decay into account in the above equation.
Thus, from the amplitude of the oscillation of $p_{L0}$ we could extracted the $u/h=$12.3(4) Hz, yielding a ratio between interactions of $U_{\uparrow\downarrow}/U=$ 99.0(1)\%.

\section{Numerics}
To compare with theoretical models, we numerical simulate the two-component BHM and spin-1 Heisenberg model. We adapted the two-component BHM from the BHM in \textit{TeNPy} package \cite{Hauschild2018}. For the ground state (zero-temperature) results we used the density matrix renormalization group (DMRG) method with a bond dimension of 30. For the dynamics simulation, we used the time evolving block decimation (TEBD) method with a bond dimension of 200. We also simulate the effective spin-1 Heisenberg model. Their results agree with each other. For the finite-temperature simulation, we exploited the purification matrix product state (MPS) implemented in the \textit{TeNPy}. In this case, it is intractable for calculation with the two-component BHM thus we only obtained the finite temperature results of the Heisenberg model. The bond dimension for the finite-temperature simulation is also 200.

\FloatBarrier
\renewcommand{\addcontentsline}[3]{}
\bibliographystyle{naturemag}

\end{document}